\documentclass[
 reprint,
superscriptaddress,
 amsmath,amssymb,
 aps,
]{revtex4-2}

\usepackage{graphicx}
\usepackage{dcolumn}
\usepackage{bm}
\usepackage{mathtools} 
\usepackage[hidelinks]{hyperref}
\usepackage{xcolor}
\usepackage{booktabs}

\makeatletter
\def\maketitle{
\@author@finish
\title@column\titleblock@produce
\suppressfloats[t]}
\makeatother

\usepackage[nottoc,notbib]{tocbibind}

\usepackage{enumerate}
\newcommand{\R}{\mathbb{R}}		

\begin{document}

\preprint{APS/123-QED}

\title{Experimental Kuramoto Platform}

\author{Leandro~Freitas}
\email{leandrofabreu@ufmg.br}
\affiliation{%
 Department of Industrial Automation and Information Technology, Instituto Federal de Educa\c{c}\~ao, Ci\^encia e Tecnologia de Minas Gerais, Campus Betim, Betim, MG 32677-562, Brazil
 }%
\affiliation{%
 Department of Electronic Engineering, Universidade Federal de Minas Gerais, Belo Horizonte, MG 31270-901, Brazil
}%

\author{Victor~H.~S.~Bittencourt}
\affiliation{%
 Department of Industrial Automation and Information Technology, Instituto Federal de Educa\c{c}\~ao, Ci\^encia e Tecnologia de Minas Gerais, Campus Betim, Betim, MG 32677-562, Brazil
 }%
\affiliation{%
 Graduate Program in Electrical Engineering, Universidade Federal de Minas Gerais, Belo Horizonte, MG 31270-901, Brazil
}%
\author{Débora~D.~Superbi}
\author{Wanderley~C.~Silva~Junior}
\affiliation{%
 Department of Industrial Automation and Information Technology, Instituto Federal de Educa\c{c}\~ao, Ci\^encia e Tecnologia de Minas Gerais, Campus Betim, Betim, MG 32677-562, Brazil
 }%

\author{Arthur~N.~Montanari}
\affiliation{
 Center for Network Dynamics, Northwestern University, Evanston, IL 60208, USA
}
\affiliation{
 Department of Physics and Astronomy, Northwestern University, Evanston, IL 60208, USA
}%

\author{Luis~A.~Aguirre}
\affiliation{%
 Department of Electronic Engineering, Universidade Federal de Minas Gerais, Belo Horizonte, MG 31270-901, Brazil
}%

\date{\today}

\begin{abstract}
Phase reduction provides a principled relationship between high-dimensional oscillator dynamics and low-dimensional phase models such as the Kuramoto model. However, connecting analytical results derived from the Kuramoto framework to experimental data remains an open challenge, as many physical platforms violate the weak-coupling assumptions underlying standard phase reduction.
Here, we present an experimental platform based on electronic quadrature oscillators whose phase dynamics are, under certain assumptions, equivalent to the Kuramoto model. Crucially, this equivalence is achieved for both weak and strong coupling through a non-diffusive coupling circuit, and the design approach supports arbitrary (weighted and directed) network topologies while preserving scalability and oscillation regularity. Across different coupling schemes, we demonstrate that the platform reliably reproduces phase transitions predicted by the Kuramoto model for both global and cluster synchronization. This cost-effective platform enables systematic experimental validation of analytical and numerical results for synchronization in complex networks.
\end{abstract}

\maketitle

\textit{Introduction}\textemdash
The Kuramoto model \cite{kur/75} has become the most paradigmatic framework for studying synchronization phenomena in complex network systems~\cite{rod_eal/16}. A major reason for its prominence is that this model arises universally from a phase reduction of smooth, autonomous limit-cycle oscillators \cite{nak/16}. 
Near a stable periodic orbit, the high-dimensional dynamics of each such oscillator can be approximated by the evolution of a single phase variable $\phi_i\in\mathbb S$. For a network of $N$ coupled limit-cycle oscillators, this reduction yields the following phase dynamics:
\begin{equation}
\dot{\phi}_i = \omega_i + K \sum^{N}_{j=1} A_{ij} \sin(\phi_j-\phi_i), \quad \text{for} \,\, i = 1,\ldots, N. \label{eq:kuramoto}
\end{equation}
\noindent
In the regime of weak coupling strength $K$, the interaction between oscillators manifests as phase-phase coupling functions, of which the sinusoidal term represents the leading harmonic.
Despite its simplicity, the Kuramoto model exhibits a rich repertoire of dynamical behaviors \cite{Montbrio2004,zhang2017incoherence,menara2019stability,Montanari2019} and phase transitions \cite{gomez2011explosive,pruser2024,xu2025} that depend explicitly on the underlying network structure, encoded by the adjacency matrix $A\in\R^{N\times N}$, and the heterogeneity among oscillators, encoded by the natural frequency $\omega_i$.

A key advantage of this phase-reduction approach is that it facilitates theoretical and computational analysis by eliminating the influence of phase-amplitude coupling, higher-order harmonic interactions, and other additional degrees of freedom. However, a notable shortcoming is that physical systems, more often than not, operate in regimes that cannot be accurately modeled by the Kuramoto approximation. These include systems subject to strong coupling \cite{dewanjee2024optimal}, pronounced amplitude fluctuations \cite{kotani2020nonlinear,nijholt2022emergent}, or non-equilibrium dynamics \cite{bressloff2020phase}, all of which invalidate the underlying assumptions for phase reduction. Hence, there is a gap between the Kuramoto model and the behavior observed in experimental setups designed to probe such systems,
constraining the applicability of Kuramoto-based analysis to narrow regions of experimental parameter space.
An experimental platform specifically designed to reproduce the Kuramoto model, without invoking the conventional phase-reduction assumption, would be a welcome asset to encompass a larger set of literature-relevant conditions.

\begin{figure*}[t]
    \includegraphics[width=0.87\textwidth]{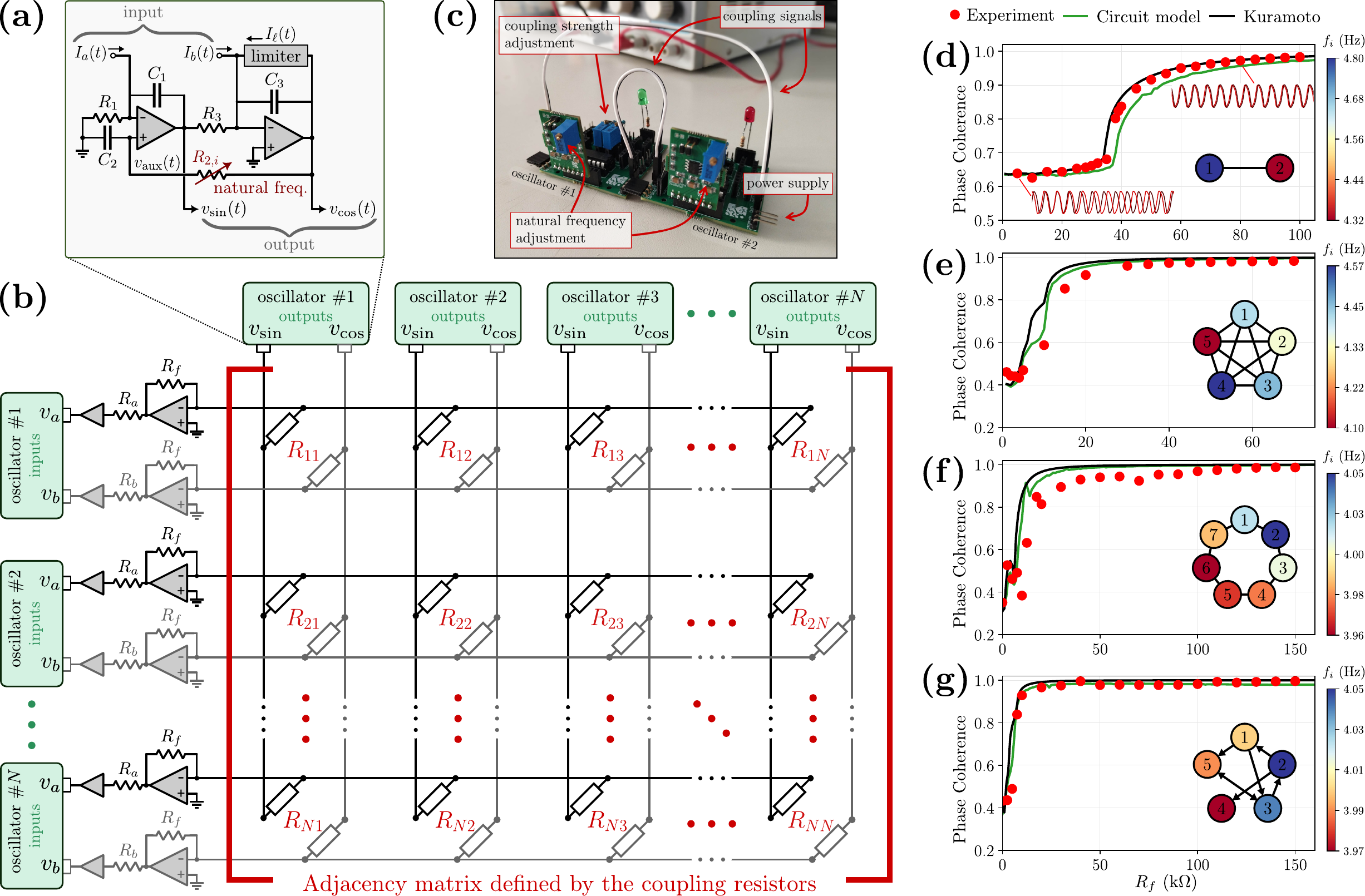}
    \caption{Experimental platform of the Kuramoto model. 
    (a)~Circuit diagram of a quadrature oscillator \cite{Graeme1971} with two input ($I_a, I_b$) and two output ($v_{\sin}, v_{\cos}$) signals. The oscillator natural frequency $\omega_i$ is tuned via the potentiometer $R_{2,i}$. The complete circuit diagram is shown in SM \cite{sm}, Fig.~S1.
    (b)~Coupling circuit implementing the adjacency matrix, with entries defined by the coupling resistors $R_{ij}$. The resistors $R_f$ control the global coupling strength $K$.
    (c)~Printed circuit board (PCB) of a pair of coupled quadrature oscillator.
    (d--g)~Experimental validation of the synchronization transition as a function of the coupling strength, shown for different network structures: (d)~symmetric pair, (e)~all-to-all, (f)~ring, and (g)~directed.
    The node colors encode the heterogeneity in $\omega_i$ (specific values are listed in the SM \cite{sm}, Table~SI).
    Solid lines show numerical simulations from the Kuramoto (black) and circuit (green) ODE model, while red data points indicate the phase coherence measured from the experimental time series.
    For illustration, the recorded time series of the signals $v_{\sin,i}(t)$ are shown in the inset of panel d.
    \label{fig:mainFigure}}
\end{figure*}

New scientific theories often advance through a cycle of i) observations that inspire hypotheses and mathematical models, ii) experiments in controlled environments that isolate the phenomena of interest, and iii) analyzes that refine or extend the initial ideas. In oscillator networks, approaches to account for empirical features\textemdash ranging from delayed signal propagation and time-varying interaction patterns to stochastic perturbations\textemdash have led to numerous extensions of the Kuramoto model \cite{Schuster1989,Crawford1999,Gutirrez2011,Rosario2015,yue2020model,lizier2023analytic,Mercado-Uribe2025}.
However, experimental studies have traditionally proceeded in the opposite direction: physical systems are built first and subsequently shown to be \textit{approximately} described by Kuramoto-like equations~\cite{kiss2002emerging,Pisarchik2009,Ochs2019,nijholt2022emergent,Pandey2024,pando2024synchronization}. These approximations typically hold only for weak coupling, narrow parameter configurations, or regular network structures. Hence, a controlled experimental environment for the study of the Kuramoto model is still lacking. Among these efforts, electronic  circuits based on coupled Wien bridge oscillators \cite{English2015,English2016,Aravind2024} have successfully reproduced Kuramoto dynamics under all-to-all coupling, but their designs cannot be readily generalized to arbitrary or programmable network topologies.

In this Letter, we address these limitations by introducing a novel electronic implementation of the Kuramoto model built on the well-established quadrature circuit and equipped with a specially designed coupling architecture that supports arbitrary, user-defined network topologies. Crucially, we show that, with a non-diffusive coupling scheme, the circuit equations can be well approximated by the Kuramoto model without imposing the weak-coupling assumption that underlies many of the commonly used phase-reduction approaches. As a result, the coupling weights, network topology, and oscillator heterogeneity can be directly adjusted through the tuning of low-cost analog circuit components. We validate this correspondence by comparing simulations of Kuramoto dynamics with measurements from experimentally built networks. Our results show that even for large frequency mismatches and strong coupling, the proposed experimental platform can accurately capture a variety of dynamical behaviors in the Kuramoto model, including phase transitions and cluster synchronization.

\smallskip
\textit{Quadrature oscillator circuit}\textemdash
The Kuramoto model assumes that each uncoupled oscillator evolves on a stable, single-frequency limit cycle of fixed amplitude such that its dynamics can be reduced to a single phase variable. The quadrature circuit in Fig.~\ref{fig:mainFigure}(a) is a natural, low-cost hardware implementation of an individual uncoupled oscillator which we show to be closely connected to the Kuramoto model. The voltage dynamics at the output nodes follows directly from Kirchhoff’s laws, and therefore each oscillator is described by
\begin{align}
\dot{v}_{\sin,i} &=  \frac{1}{\tau_{2,i}} v_{\cos,i} + \left( \frac{1}{\tau_1} - \frac{1}{\tau_{2,i}} \right) v_{\mathrm{aux},i} - \frac{1}{C_1} I_{a,i} , \nonumber \\
\dot{v}_{\cos,i} &= - \frac{1}{\tau_3} v_{\sin,i} - \frac{1}{C_3} I_{\ell,i} - \frac{1}{C_3} I_{b,i}, \nonumber \\ 
\dot{v}_{\mathrm{aux},i} &= \frac{1}{\tau_{2,i}} v_{\cos,i} - \frac{1}{\tau_{2,i}} v_{\mathrm{aux},i}, \label{eq:quad_classic}
\end{align}
where $\tau_1 = R_1C_1$, $\tau_{2,i} = R_{2,i}C_{2}$, and $\tau_3 = R_3C_3$ are the circuit parameters of the oscillator $i$,  $I_{\ell,i}(v_{\cos,i})$ is the nonlinear amplitude-stabilizing feedback current implemented through a limiter circuit and $I_{a,i}(t)$ and $I_{b,i}(t)$ denote input currents from external sources. See Supplementary Material (SM) \cite{sm}, Sec.~SI, for the derivation and analysis of the model.

The linear subsystem of~Eq.~\eqref{eq:quad_classic}, specified by $\dot{v}_{\sin,i} =  \frac{1}{\tau_{2,i}} v_{\cos,i}$ and $\dot{v}_{\cos,i} = - \frac{1}{\tau_3} v_{\sin,i}$, forms a second-order harmonic oscillator implemented in quadrature: $v_{\sin,i}(t)$ and $v_{\cos,i}(t)$ represent internal state variables shifted by $90^\circ$. With the nonlinear current inactive ($I_{\ell,i} = 0$) and no external sources ($I_{a,i}=I_{b,i} =0$), the circuit reduces to two coupled RC-integrator loops that generate a pure sinusoid with frequency
\begin{equation}
    \tilde\omega_i \approx 1/\sqrt{\tau_{2,i}\tau_3},
\label{eq.naturalfreq}
\end{equation}
which is set independently by the resistor $R_{2,i}$ of each oscillator $i$  \cite{Graeme1971}.

In practice, the circuit parameters are chosen so that the linear subsystem unstable. The nonlinear current $I_{\ell,i}$ provides amplitude regulation, converting the unstable linear subsystem into a robust limit cycle oscillator. When $|v_{\cos,i}(t)|\geq V_{\rm max}$, the limiter circuit increases $|I_{\ell, i}|$, acting as a weak nonlinear damper (SM~\cite{sm}, Fig.~S2). In practice, $v_{\sin,i}(t)\approx \sin(\tilde\omega_i t)$ (exhibiting a nearly ideal sinusoidal waveform), while $v_{\cos,i}(t)$ may show mild distortion due to nonlinear damping since its node is directly connected to the limiter (SM~\cite{sm}, Fig.~S5). Consequently, the evolution of an uncoupled quadrature oscillator is effectively captured by $v_{\sin,i}(t)$ and can be described by the phase equation $\dot\theta_i = \tilde\omega_i$.

\smallskip
\textit{Experimental Kuramoto platform}\textemdash
The quadrature oscillator, in its traditional form, is an autonomous circuit without external inputs, commonly used as a compact, low-cost sinusoidal signal generator. To realize an experimental Kuramoto platform, we i) introduce two external signals, $I_{a,i}(t)$ and $I_{b,i}(t)$, to each quadrature oscillator circuit and ii) design a coupling architecture that allows the inherently 3-dimensional quadrature oscillators to map directly onto the sinusoidal phase coupling of the 1-dimensional Kuramoto oscillator.

A natural approach is to interconnect oscillators through diffusive coupling \cite{Pecora1990,Heagy1994,Montanari2020,Baziliauskas2006}, which is widely used in synchronization studies and compatible with analytical tools such as the master stability function~\cite{pecora1998,nishikawa2006synchronization,pecora2014cluster}. Indeed, the Kuramoto model \eqref{eq:kuramoto} is itself diffusive as each pairwise coupling $(i,j)$ vanishes when $\phi_i = \phi_j$. However, implementing diffusive coupling directly on a quadrature oscillator turns out to be insufficient: a single-terminal diffusive coupling cannot generate a sinusoidal coupling once the circuit dynamics are reduced to its phase. To illustrate this, consider an oscillator $i$ coupled to an oscillator $j$ through the input $I_{a,i} = 0$ and $I_{b,i} = - v_{\sin,j}/R_3$, so that
\begin{equation}
\begin{aligned}
    \dot v_{\cos,i} &= \frac{1}{\tau_3} \left(v_{\sin,j} - v_{\sin,i}\right) - \frac{1}{C_3} I_{\ell,i} \\
                    &= \frac{1}{\tau_3} \left(r_j \sin\phi_j - r_i\sin\phi_i\right) - \frac{1}{C_3} I_{\ell,i},
\end{aligned}
\end{equation}
where we have applied the polar transformation $v_{\sin,i} = r_i\sin\phi_i$ and $v_{\cos,i} = r_i \cos\phi_i$.  This coupling acts diffusively on voltage levels, but does not produce the phase coupling $\sin(\phi_j-\phi_i)$ required by the Kuramoto model.

To realize the correct phase coupling for a network of $N$ oscillators, we introduce a \textit{dual, non-diffusive} coupling architecture in which each oscillator receives the external currents: 
\begin{equation}
\begin{aligned}
 I_{a,i} &= - \frac{R_f}{R_a} \sum_{j=1}^N A_{ij} v_{\sin,j} - \frac{v_{\mathrm{aux},i}}{R_a}, \\
 I_{b,i} &= - \frac{R_f}{R_b} \sum_{j=1}^N A_{ij} v_{\cos,j}.
\label{eq:va_vb_n2}
\end{aligned}
\end{equation}
\noindent
These signals directly inject weighted combinations of the quadrature voltages of the coupled oscillators through the summing circuits implemented by the resistor network in Fig.~\ref{fig:mainFigure}(b).
Here, $A_{ij} := R_{ij}^{-1}$ defines the network adjacency matrix, with each interaction strength independently adjusted by a potentiometer, and $R_f$ determines the global coupling strength. Disconnecting a potentiometer ($R_{ij}\rightarrow \infty$) sets $A_{ij} \to 0$. Notably, this resistor network can be asymmetric, allowing for the implementation of directed networks with heterogeneous weights. Because each directed coupling $j \rightarrow i$ uses two terminals ($I_{a,i}$ and $I_{b,i}$), a bidirectional coupling $i\rightleftarrows j$ is achieved with four-wire connections [Fig.~\ref{fig:mainFigure}(c)].

\smallskip
\textit{Mapping the circuit and Kuramoto models}\textemdash
Using the polar representation $v_{\sin,i} = r_i\sin\phi_i$ and $v_{\cos,i} = r_i \cos\phi_i$, we express Eqs.~\eqref{eq:quad_classic} and \eqref{eq:va_vb_n2} in the coordinates $(\phi_i,r_i,v_{\mathrm{aux},i})$. 
After trigonometric manipulations (SM~\cite{sm}, Sec.~SII), we obtain the following system:
\begin{align}
\dot{\phi}_i &= \frac{\cos^2(\phi_i)}{\tau_{2,i}} + \frac{\sin^2(\phi_i)}{\tau_3} + \frac{R_f}{\tau_a} \sum_{j=1}^N A_{ij}\frac{r_j}{r_i} \sin(\phi_j - \phi_i)  \nonumber\\ &+ \left( \frac{1}{\tau_1} - \frac{1}{\tau_{2,i}} + \frac{1}{\tau_a} \right) \frac{\cos(\phi_i)}{r_i} v_{\mathrm{aux},i}  + \frac{1}{C_3 r_i} \sin(\phi_i) I_{\ell,i}, \nonumber\\
\dot{r}_i &= \left( \frac{1}{\tau_{2,i}} - \frac{1}{\tau_3} \right) r_i \sin(\phi_i) \cos(\phi_i)
  \nonumber\\ &+ \frac{R_f}{ \tau_a} 
  \sum_{j=1}^N A_{ij}r_j\cos(\phi_j-\phi_i)
   \nonumber\\ & + \left( \frac{1}{\tau_1} - \frac{1}{\tau_{2,i}} + \frac{1}{\tau_a} \right) \sin(\phi_i) v_{\mathrm{aux},i}
  -  \frac{1}{C_3} \cos(\phi_i) I_{\ell,i}
  , \nonumber\\
\dot{v}_{\mathrm{aux},i} &=  \frac{1}{\tau_{2,i}}  r_i \cos(\phi_i) -  \frac{1}{\tau_{2,i}}  v_{\mathrm{aux},i} , \label{eq:dphi_dt_coup_text}
\end{align}
where the component values satisfy $R_a C_1 = R_b C_3 \eqqcolon \tau_a $.
To align the quadrature oscillator with the Kuramoto model, we impose two design conditions: i) $\tau_{2,i} = \tau_3$, ensuring symmetric quadrature dynamics, and ii) $\tau_1^{-1} = \tau_{2,i}^{-1} - \tau_a^{-1}$, which cancels additional phase-voltage coupling terms.
Under these conditions, the phase dynamics in Eq.~\eqref{eq:dphi_dt_coup_text} reduces to
\begin{equation}
\dot{\phi}_i = \tilde\omega_i + K \sum^{N}_{j=1} A_{ij} \frac{r_j}{r_i} \sin(\phi_j-\phi_i) + \frac{1}{C_3 r_i} \sin(\phi_i) I_{\ell,i}, 
\label{eq:quad_kuramoto}
\end{equation}
where $\tilde\omega_i = 1 / {\tau_{2,i}}$, $K = R_f/{\tau_a}$, and $A_{ij}  = R_{ij}^{-1}$ define the mapping between the Kuramoto model and the quadrature circuit.
Eq.~\eqref{eq:quad_kuramoto} differs from the Kuramoto model \eqref{eq:kuramoto} in only two respects: i) the presence of phase-amplitude coupling through the factor $r_j/r_i$ and ii) the nonlinear limiter current. In practice, the current limiter $I_{\ell,i}$ tightly regulates the amplitude signals $r_i$ against external disturbances, ensuring $r_i\approx r_j$ across oscillators. Moreover, the contribution of $I_{\ell,i}$ to the phase dynamics is periodic and can therefore be absorbed into an effective natural frequency. As a result, we have that Eq.~\eqref{eq:quad_kuramoto} can be approximately reduced to Eq.~\eqref{eq:kuramoto} by defining $\omega_i\approx \tilde\omega_i + \langle \frac{1}{C_3r_i} \sin(\phi_i) I_{\ell,i} \rangle$, where $\langle \cdot\rangle$ denotes a time average over a single period. Experimentally, $\omega_i$ can be independently tuned using the potentiometer $R_{2,i}$ and measured with an oscilloscope. These practical considerations are further discussed and supported by numerical evidence in the SM \cite{sm}, Sec.~SIII.

\smallskip
\textit{Experiments on phase transitions}\textemdash
We now validate the capability of our model to reproduce the dynamical behavior and phase transitions of a Kuramoto model. Experimental details on component, PCB design, and methodology are reported in the SM~\cite{sm}, Secs.~SIV and SV.
Fig.~\ref{fig:mainFigure}(d)--(g) presents four experiments designed to test whether the experimental platform could reproduce the synchronization transition of a Kuramoto model for different network structures and a wide range of coupling strengths. In each experiment, the adjacency matrix $A$ was fixed, and the state variables $v_{\sin,i}(t)$ were measured to calculate the individual phase (via the Hilbert transform). The experimental results were compared with numerical simulations obtained from the Kuramoto model \eqref{eq:kuramoto} and the coupled quadrature oscillator model [Eqs.~\eqref{eq:quad_classic} and \eqref{eq:va_vb_n2}], using consistent parameter values.

Across all cases, the resulting order parameter $R = \langle | \frac{1}{N}\sum_j e^{{\rm i}\phi_j(t)} | \rangle$ shows strong agreement among experimental measurements, numerical simulations of the circuit model, and Kuramoto model predictions. Quantitatively, the mean relative deviation between the Kuramoto model and the numerical simulations of the coupled oscillator circuit lies in the range 2.99--6.63\%, while the deviation between the Kuramoto model and the experimental data falls within 1.21--8.83\% across the explored coupling strengths. These discrepancies are naturally attributable to experimental imperfections, including component tolerances and measurement noise. The same performance was verified in stronger coupling regimes ($K \approx 18$, with $R_f=400~\textrm{k}\Omega$) within the fully connected network. Considering the network size, this coupling strength equates to $30$ times the maximum natural frequency spread (see SM~\cite{sm}, Sec.~SV for further discussion).

Among the tested configurations, the two-oscillator system and the directed network topology exhibit the closest agreement between theory, numerics, and experiments, whereas the ring topology shows the largest mismatch between numerical and experimental results. Nevertheless, a supplementary tolerance analysis for the ring network demonstrates that the experimental measurements fall within the expected standard deviation once component variability is taken into account (SM \cite{sm}, Fig.~S7). 
The accuracy of the Kuramoto model in the directed network topology is particularly remarkable given that directed networks are generally more sensitive to weighted perturbations (i.e., small discrepancies in link weights between the experimental and numerical setups) \cite{nishikawa2017sensitive}. This robustness indicates that the synchronization transition captured by the Kuramoto description remains stable under realistic experimental imperfections, even in network configurations where asymmetry and directionality amplify the effects of parameter variations.

\begin{figure*}[th]
    \includegraphics[width=0.88\textwidth]{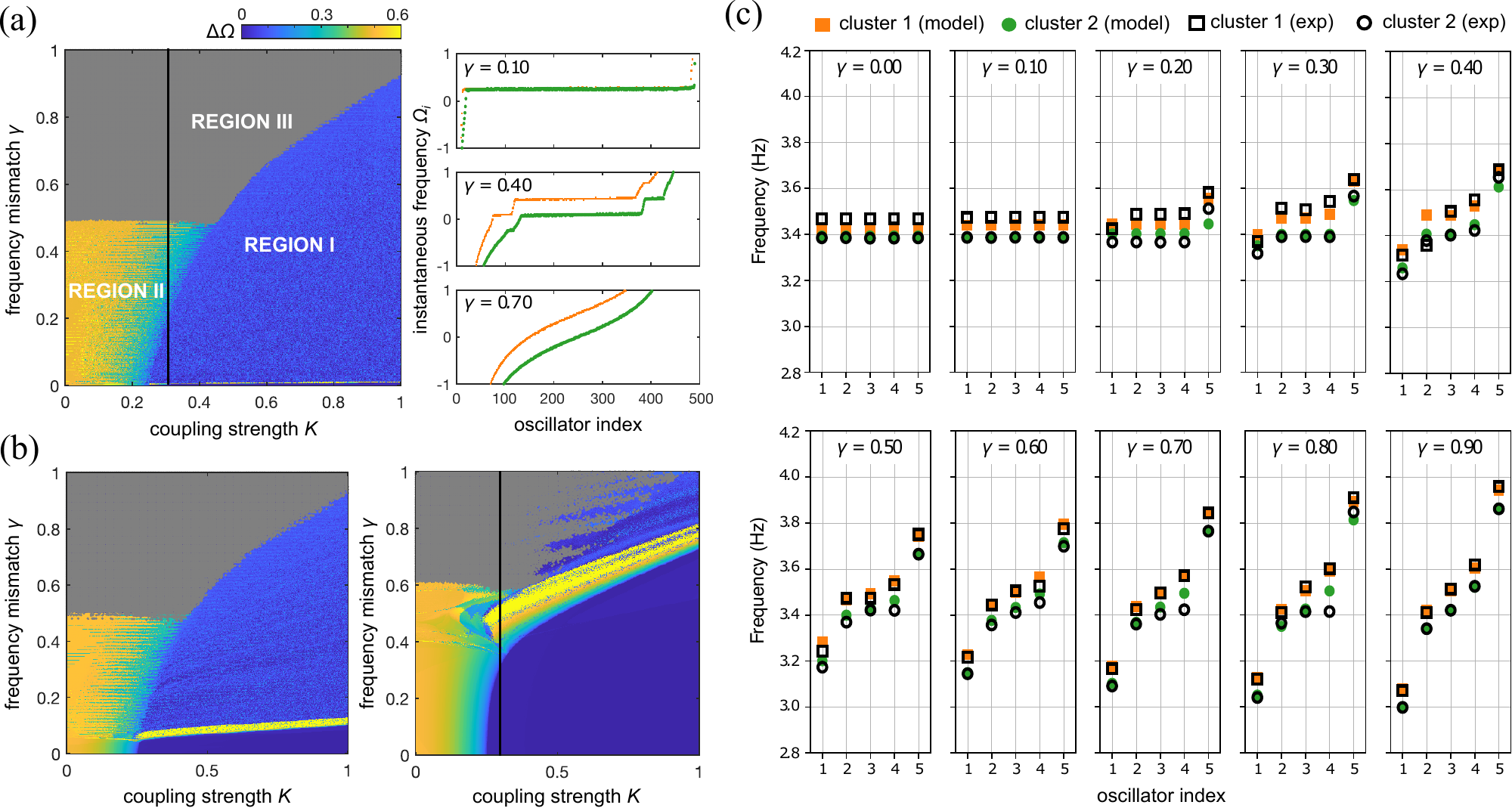}
    \caption{ 
    Cluster synchronization in the experimental Kuramoto platform.
    (a) $(K,\gamma)$-phase diagram in a weakly coupled network of two clusters, each with $M=500$ Kuramoto oscillators. Color bar represents the instantaneous frequency mismatch $\Delta\Omega$ between the two clusters. Regions I, II, and III correspond respectively to global synchronization (blue region, where $\Delta\Omega \approx 0$), cluster synchronization (light blue to yellow region, where $\Delta\Omega > 0$), and desynchronization (gray region). 
    The right panels show the instantaneous frequencies $\Omega_i$ of individual oscillators, color-coded by cluster, for representative values of $\gamma$ and fixed $K=0.3$.
    (b) $(K,\gamma)$-phase diagram for smaller cluster sizes: $M=50$ (left) and $M=5$ (right).
    (c) Experimental validation of the $(K,\gamma)$-phase diagram for the case $M=5$, using a fixed $K=0.3$ and varying $\gamma$. Filled markers indicate numerical results from the Kuramoto model, while outlined markers indicate experimental measurements.
    \label{fig.cluster}}
\end{figure*}

\smallskip
\textit{Experiments on cluster synchronization}\textemdash Beyond complete synchronization, we demonstrate that our platform can also reproduce richer spatiotemporal collective behavior such as cluster synchronization. To this end, we consider two populations (clusters) of $M$ Kuramoto oscillators coupled asymmetrically as follows:
\begin{align}
\dot{\theta}_i^{(1,2)} &= \omega_i^{(1,2)} - \frac{K_p}{M} \sum^{M}_{j=1} \sin\left(\theta_i^{(1,2)}-\theta_j^{(1,2)}\right) \nonumber\\
& - \frac{K}{M} \sum^{M}_{j=1} \sin\left(\theta_i^{(2,1)}-\theta_j^{(1,2)}\right) , \label{eq:montbrio_oscillators}
\end{align}
where $\theta_i^{(1,2)}$ is the phase of the $i$th oscillator in the population $1$ or $2$. The parameters $K_p$ and $K$ represent the intra-cluster and inter-cluster coupling strengths, respectively. The natural frequencies are chosen according to a Lorentzian distribution as
\begin{equation}
    g^{(1,2)}(\omega) = \left(\frac{\gamma}{\pi}\right) \left[\gamma^2 + \left(\omega - \bar{\omega}^{(1,2)}\right)^2\right]^{-1}
\label{eq:lorentzian}
\end{equation}
where $\bar{\omega}^{(1,2)}$ sets the center frequency of each population and $\gamma$ controls the width of the distribution, quantifying the degree of frequency heterogeneity. At the thermodynamic limit $M\to\infty$, each population can be described by the continuous phase density function $\rho^{(1,2)}(\theta,t,\omega)$ that represents the fraction of oscillators with natural frequency $\omega$ and phase $\theta$ at time $t$ \cite{Montbrio2004}.

Fixing $K_p=1$, three distinct dynamical regimes emerge in the $(K,\gamma)$-phase diagram: global synchronization (region I), cluster synchronization (region II), and complete desynchronization (region III). Figure~\ref{fig.cluster}(a) depicts these regions as obtained from numerical simulations with $M = 500$, for which the system behavior closely approximates the thermodynamic limit. In the synchronized regimes (regions I and II), oscillators are frequency locked and organize into discrete plateaus corresponding to Shapiro steps, which provide a clear signature of phase synchronization [Fig.~\ref{fig.cluster}(a, right panels)]. As $M$ reduces, finite-size effects become increasingly pronounced and the continuum assumption underlying the thermodynamic description gradually breaks down. Consequently, the boundaries between dynamical regimes deviate progressively from the thermodynamic-limit phase diagram, as illustrated in Fig.~\ref{fig.cluster}(b).

The proposed platform was used to experimentally validate these results for a finite system with $N=2M=10$ oscillators. Figure~\ref{fig.cluster}(c) shows the experimental outcomes across a range of $\gamma$ values (see SM \cite{sm}, Fig.~S8, for the complete set of experiments). The observed behavior closely matches analytical predictions. For small $\gamma$, both populations exhibit global synchronization (region I), while cluster synchronization emerges within each population for $\gamma \gtrsim 0.2$, marking the transition to region II. As $\gamma$ is further increased ($\gamma \gtrsim 0.6$), intra-population cluster synchronization dissolves and the system transitions to complete desynchronization (region III), with each oscillator exhibiting distinct dynamics.
Overall, these experiments demonstrate the robustness of the analytical results reported in Ref.~\cite{Montbrio2004} under realistic experimental conditions, including noise, parameter variability, and other nonideal effects inherent to physical implementations.

\textit{Conclusions}\textemdash
This work presents a low-cost and highly accessible experimental platform for validating analytical and numerical studies of coupled Kuramoto oscillators with arbitrary network topologies. The mathematical correspondence between the proposed circuit and the Kuramoto model on heterogeneous networks is established for both weak and strong coupling regimes, without relying on conventional phase-reduction  approximations \cite{nak/16}. As demonstrated through the experimental validation of phase transitions in complete and cluster synchronization regimes, the proposed circuit provides a flexible platform for investigating a wide range of behaviors exhibited by the Kuramoto model.

Future work may extend this platform to incorporate more complex network dynamics. For example, the phase delay introduced in the Kuramoto–Sakaguchi model can be readily implemented by adding capacitive impedance to the coupling circuit. This extension is particularly relevant for the experimental investigation of chimera states and crowd synchronization in coupled population models \cite{panaggio2015chimera}, which to date have largely been confined to complex and high-cost experimental platforms involving chemical \cite{tinsley2012chimera}, optical \cite{hagerstrom2012experimental,mahler2020experimental}, mechanical \cite{martens2013chimera}, or photoelectrochemical \cite{schmidt2014coexistence} oscillators.
Similarly, signed networks can be implemented using inverter circuits, while adaptive networks can be realized through digital switches that dynamically modify network topology and connectivity. Digital potentiometers can further enable real-time adaptation of natural frequencies or coupling weights for feedback control. Overall, this platform offers a promising experimental testbed for studying disordered systems \cite{Aravind2024,ocampo2025frequency,barioni2025interpretable}, physical computing \cite{mallick2020using,brown2025charge,allibhoy2025global}, and collective dynamics in higher-order networks \cite{nijholt2022emergent,battiston2025collective}.

\textit{Acknowledgments}\textemdash This research was supported by the Brazilian agencies Fundação de Amparo à Pesquisa do Estado de Minas Gerais - FAPEMIG (Grant APQ-00781-21), Conselho Nacional de Desenvolvimento Científico e Tecnológico - CNPq (Grant 409487/2021-0), and Coordenação de Aperfeiçoamento de Pessoal de Nível Superior - CAPES (finance code 001). LAA also acknowledges CNPq (Grant 305115/2024-3).

\textit{Data availability}\textemdash The code and data supporting the findings of this article are openly available \cite{data_aval}. The PCB design is also publicly available through this repository.

\let\oldaddcontentsline\addcontentsline
\renewcommand{\addcontentsline}[3]{}


\providecommand{\noopsort}[1]{}\providecommand{\singleletter}[1]{#1}%

\end{document}